\documentclass[preprint,12pt]{elsarticle}

\usepackage{amssymb}
\usepackage{macros_gildas}
\usepackage{macros_rolando}
\usepackage{macros}

\usepackage{pifont}

\usepackage{subfigure}

\usepackage{url}

\usepackage[pdfpagelabels]{hyperref}
\hypersetup{colorlinks=true, linkcolor=blue, citecolor=blue, urlcolor=blue}

\newtheorem{definition}{Definition}

\begin{document}

\begin{frontmatter}

\title{Comparing Distance Bounding Protocols: a Critical Mission Supported
by Decision Theory}


\author[renne,belgium]{Gildas Avoine}

\author[snt]{Sjouke Mauw}

\author[snt]{Rolando Trujillo-Rasua \corref{cor1}}

\address[renne]{INSA Rennes, IRISA UMR 6074, Institut Universitaire 
de France}

\address[belgium]{Universit\'{e} catholique de Louvain, Belgium}

\address[snt]{Interdisciplinary Centre for Security, Reliability and Trust \\
University of Luxembourg}

\cortext[cor1]{Corresponding author. Phone: +352 466 644 5458. Fax: 466 644 3 
5458. Email: 
rolando.trujillo@uni.lu}

\begin{abstract}
Distance bounding protocols are security countermeasures designed to thwart 
relay attacks. Such attacks consist in relaying messages exchanged between two 
parties, making them believe they communicate directly with each other. 
Although distance bounding protocols have existed since the early nineties, 
this research topic resurrected with the deployment of contactless systems, 
against which relay attacks are particularly impactful. Given the impressive 
number of distance bounding protocols that are designed every year, it becomes 
urgent to provide researchers and engineers with a methodology to fairly 
compare the protocols in spite of their various properties. This paper 
introduces such a methodology based on concepts from the decision making field. 
The methodology allows for a multi-criteria comparison of distance bounding 
protocols, thereby identifying the most appropriate protocols once the context 
is provided. As a side effect, this paper clearly identifies the protocols that 
should no longer be considered, regardless of the considered scenario.
\end{abstract}


\end{frontmatter}


\section{Introduction}\label{section:introduction}



Distance bounding protocols are the most popular countermeasures against
relay attacks. In a relay attack on an authentication protocol,
an adversary aims to convince the verifier that he directly
communicates with the genuine prover, while the adversary is actually
in the middle and relays the messages exchanged between the two
parties. Typically, a relay attack makes the verifier believe the
prover is located within his neighborhood while he is far away.



\subsection{Relay attacks}

Conway~\cite{C2000} introduced in 1976 the concept of
a relay attack through the \emph{Chess Grandmaster problem} where a
little girl is challenged to defeat a Chess Grandmaster in
correspondence chess. The solution suggested by Conway to allow the
little girl to be successful is to perform a relay attack between two
Chess Grandmasters: the attack consequently consists in relaying the
moves received between the two Chess Grandmasters, which results for
the little girl in either a won or two draws.

Relay attacks also apply to authentication protocols as
originally proposed by Desmedt, Goutier, and Bengio at
Crypto~87~\cite{DGB1988}, whose work was later
extended by
Brassard and Quisquater in \cite{BBDGQ1991}. In their
papers, the authors refuted Shamir's claims about the Fiat-Shamir
protocol~\cite{FS1986} when he says that
the protocol is secure even when being executed one million times in a
Mafia-owned store~\cite{G1987}. Desmedt {\etal}
indeed raised that a relay attack is still possible, and they
consequently named the suggested relay attack \emph{mafia fraud}.
Since then, both terms, relay attack and mafia fraud, are used
interchangeably in the literature. Note however that Avoine
{\etal}~\cite{ABKLM2011} distinguish mafia fraud from
relay attacks by considering that the adversary cannot modify the
forwarded messages in a relay attack. This distinction allows for
representing an adversary who does not know the specifications of the
considered protocol.


Although mafia fraud was suggested late in the eighties, practical 
implementations of this type of fraud appeared much later. Mafia fraud actually 
became a real threat with the ubiquity of contactless technologies. For 
example, practical attacks were developed against Radio Frequency 
IDentification (RFID)~\cite{H2006,HK2008}, Near Field Communication 
(NFC)~\cite{FHMM2010}, and Passive Keyless Entry and Start Systems (PKES) in 
modern cars~\cite{FDC2010}. For example, off-the-shelves devices to perform 
relay attacks against PKES can be bought on Internet~\cite{bundpol}.


\subsection{Distance bounding protocols}

Mafia fraud does not rely on exploiting security protocol
vulnerabilities. Conventional security mechanisms are thus ineffective
against it. Based on an idea from Beth and Desmedt~\cite{BD1990},
Brands and Chaum suggested a countermeasure to mafia fraud that
consists in measuring the Round-Trip-Time (RTT) of 1-bit messages
exchanged between the parties, using a dedicated communication
channel~\cite{BC1993}. In their solution, the verifier measures the
round-trip time $t_m$ between the moment he sent a challenge and the
moment he receives the response from the prover.  The verifier can
consequently estimate a tight upper-bound on the distance between the
prover and the verifier by computing $d = c \cdot (t_m - t_d) / 2$,
where $c$ is the speed of light and $t_d$ is the delay induced by the
prover to compute the response, given the challenge.

Note that distance bounding protocols do not detect relay attacks in a
strict sense. Instead, they detect unexpected delays, and conclude in
such a case that a mafia fraud attack might have occurred. 
As a consequence, neither the
communication channel, nor the calculation should introduce flexible
timing during the protocol execution, since that could be exploited by
an adversary. For example, requiring the prover to perform heavy
computations in passive contactless systems may allow an adversary to
significantly reduce $t_d$ by overclocking the prover's device, which
in turn may allow the adversary to increase $t_m$ without making $d$
above the expected upper-bound. Since Desmedt {\etal}'s seminal
work~\cite{BD1990}, a conservative assumption for designing distance
bounding protocols consists in considering minimally sized messages
(typically 1-bit messages) and lightweight computations during the
time-measurement phase.

\subsection{Protocol evaluation}

Avoine {\etal} introduced in~\cite{ABKLM2011} a \emph{Framework} for
analyzing distance bounding protocols. This widely used Framework
defines four types of fraud that should be considered in the security
evaluation of distance bounding protocols. For the sake of accuracy,
the fraud definitions from~\cite{ABKLM2011} are provided
\emph{in-extenso} below.
\begin{itemize}
\item Given a distance bounding protocol, an \emph{impersonation
  fraud} attack is an attack where a lonely prover purports to be
  another one.
\item A \emph{mafia fraud} attack is an attack where an adversary
  defeats a distance bounding protocol using a man-in-the-middle
  (MITM) between the reader and an honest tag located outside the
  neighborhood.
\item Given a distance bounding protocol, a \emph{distance fraud}
  attack is an attack where a dishonest and lonely prover purports to
  be in the neighborhood of the verifier.
\item A \emph{terrorist fraud} attack is an attack where an adversary defeats
  a distance bounding protocol using a man-in-the-middle (MITM)
  between the reader and a dishonest tag located outside of the
  neighborhood, such that the latter actively helps the adversary to
  maximize her attack success probability, without giving to her any
  advantage for future attacks.
\end{itemize}
The security evaluation of a distance bounding protocol then consists
in computing the resistance of the protocol for every type of fraud, which is
done by computing the probability for an adversary to successfully
perform the considered fraud.

Since Brands and Chaum's breakthrough, many distance-bounding
protocols have been proposed\footnote{http://www.avoine.net/rfid/}, which 
deliver improvements
in terms of security (see Section~\ref{section:stateoftheart}). These proposals 
also introduce new requirements
on the protocols, e.g., to be usable on noisy channels, and
properties, e.g., to be more computationally efficient or to require
less memory. Given the various requirements and properties, a fair
methodology to compare distance bounding protocols is strongly needed.

\subsection{Contribution}

This paper introduces a methodology based on concepts from the
decision making field to perform a multi-criteria comparison of
distance bounding protocols. The methodology identifies the most
desirable protocols, given a set of required properties, and
disqualifies protocols that are dominated by better solutions whatever
the considered properties. Even though the methodology can be
understood without difficulty, applying it on a large set of distance
bounding protocols may be time-consuming. As a consequence, an
open-source computer tool was released in order to easily include
into the comparison future distance bounding protocols and new
criteria.

\section{Background}\label{section:stateoftheart}

Distance bounding protocols are authentication protocols that, in
addition, compute an upper bound on the distance between the
prover and the verifier. Since we focus on the distance bounding
properties of such protocols, we ignore any such protocol that
does not even achieve authentication, e.g., due to impersonation
attacks or key-recovery attacks~\cite{Stinson-1995-book}.
The considered
protocols are briefly introduced and classified according to their
main features, which are the features that occur most frequently in
literature and that should be taken into account to compare the protocols.
The protocols are listed in Table~\ref{table_protocol_list}.

\begin{table}
  \caption{List of protocols and their acronyms.\label{table_protocol_list}}
  \centering
  \begin{tabular}{ @{} l c  c c  @{}}
    \toprule %
    \textbf{Authors}        & \textbf{Reference} & \textbf{Year} & 
    \textbf{Acronym} \\\cmidrule(l){1-4}
    Brands and Chaum        &\cite{BC1993} & 1993 & BC \\
    \v{C}apkun, Butty\'{a}n,and Hubaux &\cite{CBH2003} & 2003 & MAD\\
    Bussard and Bagga       &\cite{BB2005} & 2005 & BB \\
    Hancke and Kuhn         &\cite{HK2005} & 2005 & HK \\
    Munilla and Peinado     &\cite{MP2008} & 2006 & MP\\
    Kim, Avoine, Koeune, Standaert, and Pereira &\cite{KAKSP2008} & 2008 & 
    Swiss-Knife \\
    Avoine and Tchamkerten  &\cite{AT2009} & 2009 & Tree-based \\
    Trujillo-Rasua, Martin, and Avoine &\cite{TMA2010} & 2010 & Poulidor \\
    Rasmussen and \v{C}apkun&\cite{RC2010} & 2010 & RC \\
    Yum, Kim, Hong and Lee   &\cite{YKHL2010} & 2010 & YKHL \\
    Kim and Avoine          &\cite{KA2011} & 2011 & KA \\
    Boureanu, Mitrokotsa, and Vaudenay &\cite{BMV2013} & 2013 & SKI \\
    Trujillo-Rasua, Martin, and Avoine &\cite{TMA2014} & 2014 & TMA \\
    \bottomrule
  \end{tabular}
\end{table}

\subsection{Compared protocols}

\subsubsection{Resistance to mafia and distance fraud.}
The earliest distance bounding protocol, introduced by Brands and
Chaum in 1993~\cite{BC1993}, consists of an initial commitment phase,
followed by $n$ rounds where the verifier sends a single-bit challenge
and receives a single-bit response from the prover. The protocol is
then completed with a final phase where the commitment is opened and a
signature of the exchanged messages is provided by the prover. The
phase during which the round trip time (RTT) is measured is known as
being the \emph{fast phase} while the other ones are known as the
\emph{slow phases}.  The BC protocol, provided in
Algorithm~\ref{algorithm:bc}, reaches the optimal security bound
$\left(1/2\right)^n$ against both mafia and distance fraud, where $n$
is the number of rounds\footnote{For every distance bounding protocol
  with a single fast phase consisting of $n$ rounds of $1$-bit
  exchanges, an adversary who answers randomly during the fast phase
  and relays all the other messages succeeds with probability
  $\left(1/2\right)^n$~\cite{ABKLM2011}.}. The authors, however, left
as an open problem the design of a distance-bounding protocol that
resists to terrorist fraud as well.

{\renewcommand{\arraystretch}{\rowh} 
  \begin{algorithm}[hbt]
    \begin{center}
      \[\pinit%
      \begin{tabular}{ccc}
        \boitegauche{\textbf{Verifier}} & \boitecentre{}  & 
        \boitedroite{\textbf{Prover}} \\
        \boitegauche{(prover's public key $K_v$)} & \boitecentre{} & 
        \boitedroite{(prover's private key $K_s$)} \\
      \end{tabular}
      \right.\]
      \interphase
      \[\pslow%
      \begin{tabular}{rcl}
        \boitegauche{}&\flechegauche{$\commit(m_1||\ldots||m_n)$}&\boitedroite{$m_i
         \in_R \{0,1\}$}\\
      \end{tabular}
      \right.\]
      \interphase
      \[\pfast%
      \begin{tabular}{rcl}
        &\textbf{begin of fast phase} &\\
        \boitegauche{Pick a random bit $c_i$} & \flechedroite{$c_i$} & 
        \boitedroite{}\\
        \boitegauche{} & \flechegauche{$r_i$}  & \boitedroite{$r_i = m_{i} 
        \oplus c_i$}\\
        &\textbf{end of fast phase}&\\
      \end{tabular}
      \right.\]
      \interphase
      \[\pslow%
      \begin{tabular}{rcl}
        \boitegauche{} & \flechegauche{$\open(\commit), \ \sign_{K_s}(c_1||r 
        _1||\ldots)$} & \boitedroite{}\\
      \end{tabular}
      \right.\]
      \[\pverif%
      \begin{tabular}{lcl}
        \boitegauche{Check $r_i$ and the RTTs\\Verify $\sign_{K_s}$} & 
        \boitecentre{}&\boitedroite{}\\
      \end{tabular}
  \right.\]
  \caption{Brands and Chaum's Protocol}\label{algorithm:bc}
    \end{center}
\end{algorithm}}


\subsubsection{Resistance to terrorist fraud.}
The challenge of designing a protocol resistant to terrorist fraud was
taken up later in 2005 by Bussard and Bagga~\cite{BB2005}, who
proposed a protocol similar in design to the BC protocol. In addition
to commitment and signature schemes, the BB protocol uses a $(2,
2)$-secret sharing scheme aimed at defeating terrorist fraud. However,
Avoine, Lauradoux, and Martin~\cite{ALM2011} demonstrated that a $(2,
2)$-secret sharing scheme is insufficient to thwart terrorist fraud: a
$(3, 3)$-secret sharing scheme should be used instead.


Later on, in 2008, a first distance-bounding protocol resistant to
some extent to terrorist fraud was suggested by Kim
{\etal}~\cite{KAKSP2008}. This protocol was named
the Swiss-knife distance-bounding protocol -- in reference to the
multi-tool Swiss army knife -- due to its ability to deal with mafia,
distance, and terrorist fraud at the same time. Nevertheless, its
resistance value of $\left(3/4\right)^n$ to both mafia and terrorist
fraud  falls far beyond the optimal security bound
$\left(1/2\right)^n$.

More recently, in 2013, the SKI family of protocols was designed by
Boureanu, Mitrokotsa, and Vaudenay~\cite{BMV2013} to counter terrorist
fraud. The SKI protocols do not perform better than existing
protocols, but they benefit from the availability of security proofs.


\subsubsection{Final slow phase and lightweight cryptographic operations.}
The boom of RFID technology in the early 21st century, impulsed by
Walmart's\footnote{Walmart is the largest retailer in the world.}
announcement of tagging pallets and cases of goods with RFID tags,
motivated Hancke and Kuhn to design the first distance-bounding
protocol for resource-constrained
devices~\cite{HK2005}. To do so, they dropped the
objective of making the protocol secure against terrorist fraud, and
focused on eliminating both the final slow phase and the need of
expensive cryptographic primitives, such as commitment and signing.
The drawback of the HK protocol is its low resistance to both distance and
mafia fraud, which is
$\left(3/4\right)^n$~\cite{HK2005,TMA2010}.


Inspired by the strengths and weaknesses of Hancke and Kuhn's
proposal, several other distance-bounding protocols were
proposed~\cite{AT2009,MP2008,KA2011,TMA2010,TMA2014,YKHL2010}.
All of them aim at improving the security to both mafia fraud and
distance fraud, while keeping the simple design of the HK protocol to make
them suitable for low-cost devices. The protocols proposed 
in~\cite{YKHL2010,TMA2014} also aim extra features such as mutual 
authentication and noise resiliency respectively.

\subsubsection{Memory.}
Among the protocols inspired by the HK protocol, the tree-based
protocol proposed by Avoine and Tchamkerten~\cite{AT2009} achieves the
best asymptotic security to mafia and distance fraud. Unfortunately,
the tree-based protocol requires an exponential amount of memory
w.r.t. the number of rounds of the fast phase. To mitigate this
problem, the authors~\cite{AT2009} suggest a trade-off between memory
requirement and security by parameterizing the depth of the tree.

Another approach by Trujillo-Rasua, Martin, and Avoine~\cite{TMA2010}
consists in using a graph instead of a tree. This protocol, named
Poulidor, requires a linear memory instead of an exponential one, but
degrades the resistance to mafia and distance fraud in comparison to
the tree-based protocol. An additional issue is that the analysis of
Poulidor is complex~\cite{T2013} and only conservative bounds on the
resistance to the various types of fraud have been provided.

To increase the resistance to mafia and distance fraud without
significantly increasing the memory requirement, Kim and
Avoine~\cite{KA2011} proceed differently and suggest a trade-off
between distance and mafia fraud resistance, which can be adapted to
any given scenario.


\subsubsection{Single-bit exchanges.}
Based on the HK protocol, Munilla and Peinado introduced a
distance-bounding protocol~\cite{MP2008} where three-state challenges
are used instead of binary challenges. This idea was later improved
and generalized by MUSE~\cite{AFM2009}, which assumes a multiple-bit
channel during the fast phase. Actually, MUSE is a technique (not a
protocol \emph{per se}) that transforms any single-bit challenge
protocol into a multiple-bit challenge protocol.  Empirical results
in~\cite{AFM2009} suggest that a MUSE transformation achieves better
security properties than the single-bit challenge counterpart. For
instance, the resistance of the BC protocol~\cite{BC1993} to mafia
fraud is $\left(1/2\right)^n$, while its MUSE transformation with a
$2$-bit channel achieves $\left(1/4\right)^n$. In both protocols, $n$
denotes the number of rounds during the fast phase, which means that
the security is measured in terms of number of rounds. However,
considering the number of bits exchanged during the fast phase,
denoted $e$, the security of both protocols becomes equal to
$\left(1/2\right)^e$. This illustrates the difficulty in comparing
protocols that require different properties concerning the channels.

\subsection{Protocol evaluation}\label{subsection:protocolevaluation}

To the best of our knowledge, Kim {\etal}~\cite{KAKSP2008} were the
first authors comparing their protocol against previously proposed
distance bounding protocols. They used a tabular form and evaluated
eight different protocols in terms of mafia and terrorist fraud
resistance, number of cryptographic operations to be performed by the
prover, noise resiliency of the protocol, privacy preservation, and
mutual authentication. In the comparisons published later on, the last
three properties are generally not considered, as done for example
in~\cite{BMV2013}. It is worth noting that the mentioned criteria are
equally important and cannot be ranked: this implies that protocols
can be compared according to one criterion at a time only. Note also
that the resistance to attacks is generally evaluated
asymptotically, i.e., when the number of rounds tends to
infinity. However, a protocol might be asymptotically better than
another protocol, while it is worse for some small number of rounds.

\begin{figure}[!htb]
  \begin{center}
    \includegraphics[width=0.65\textwidth, angle=270]{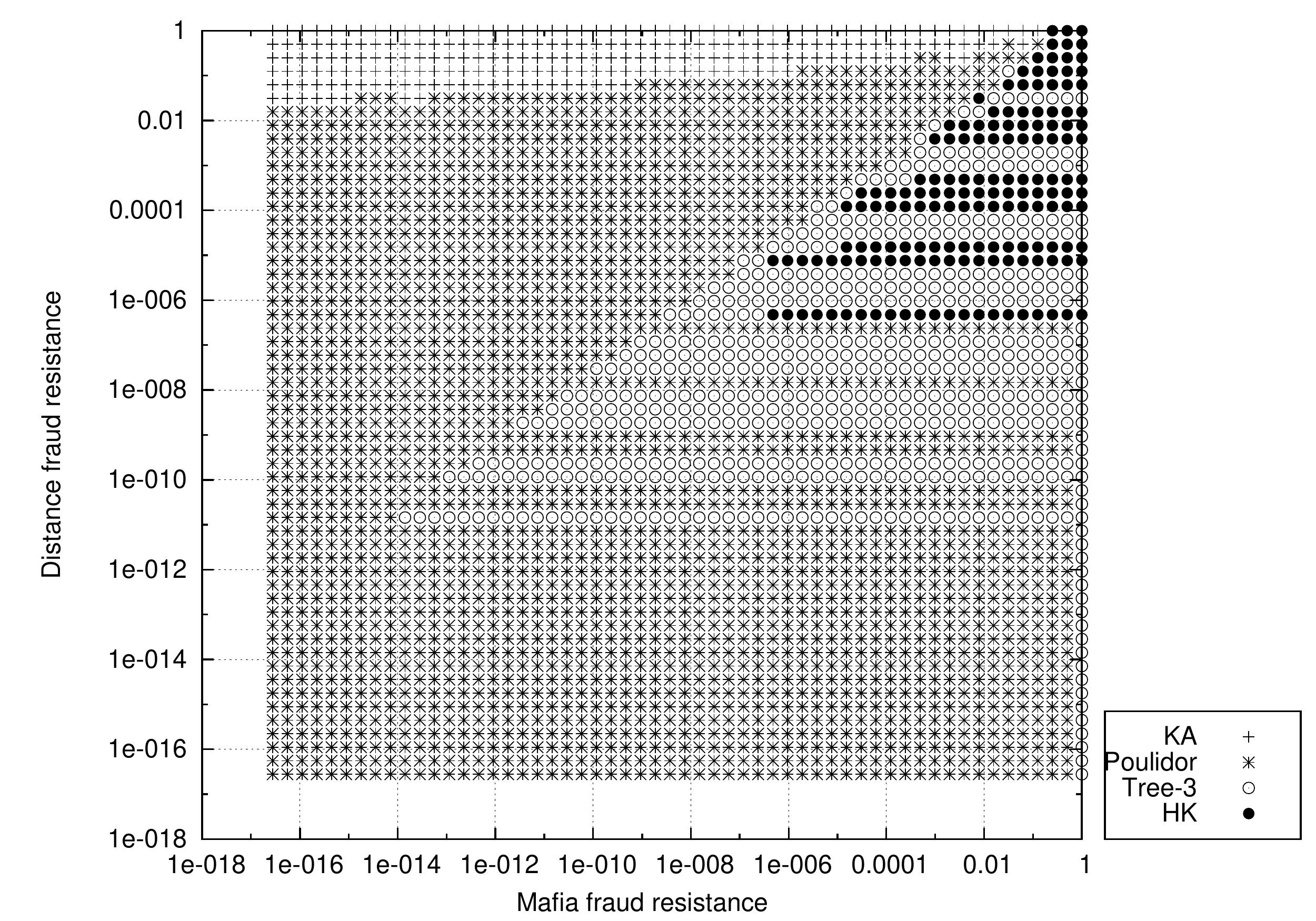}
  \end{center}
  \caption{A visual representation~\cite{TMA2010} of the comparison of
    Poulidor~\cite{TMA2010}, the HK protocol~\cite{HK2005}, the KA
    protocol~\cite{KA2009}, and the tree-based protocol using trees of
    depth $3$ (Tree-3)~\cite{AT2009}.}
  \label{fig_trade_off}
\end{figure}

Trujillo-Rasua {\etal}~\cite{TMA2010,TMA2014} suggested a significantly
different technique to compare distance bounding protocols, where the
comparison is based on two criteria and is no longer done
asymptotically. So, for every protocol and for every (discretized)
pair $(m, d)$ of mafia and distance fraud resistance values in
$[0,1]^2$,
the technique computes the minimum number of rounds $n$
needed to reach these values. For every pair $(m, d)$, the \emph{best}
protocol is the one that requires the smallest value $n$.
Figure~\ref{fig_trade_off} represents the result of the comparison
applied to the Poulidor, HK, KA, and tree-based protocols in~\cite{TMA2010}. 
The 2D
chart displays the best protocol (or one of the best protocols in case
of equality) among the four considered ones for every possible value
of mafia and distance fraud. For example, when $(m, d)=(1,1)$, the
best protocol is HK.

The comparison methodology introduced by Trujillo-Rasua {\etal} is
more advanced than the one suggested by Kim {\etal}, but its usability
remains limited. Indeed, Trujillo-Rasua {\etal}'s methodology requires
criteria that impact the objective function, which is minimizing the
value $n$. For example, applying the methodology with the criteria
``mafia fraud resistance'' and ``presence of a final slow phase'' is
meaningless, given that the presence or not of a final slow phase does
not depend on $n$. Another weakness -- although the core of the
methodology is not concerned -- is the 2D representation of the
result, which is inappropriate when considering more than two
criteria.

\section{Methodology}\label{sec:methodology}

Multi-criteria decision-making actually consists in making a decision,
namely selecting the best solution(s) in a set of possible solutions,
when the evaluation of solutions depends on several criteria. For
example, buying a car is a multi-criteria decision making problem,
because price, size, horsepower, color, etc. are different criteria
that influence the decision. Similarly, choosing a distance bounding
protocol is a multi-criteria decision-making problem where several
security and implementability criteria need to be considered. 
This
section defines the relevant \emph{attributes} that ought to be
considered in distance bounding protocols, together with the concepts of
\emph{approximate equality}, \emph{attribute spaces}, 
\emph{dominant relation}, and \emph{protocol instance}.

\subsection{Attributes}


Decision criteria are built on atomic attributes that charaterize the 
options available, namely the distance bounding protocols in our case. The 
most common attributes used in the literature to evaluate distance bounding 
protocols are related to security and implementability. These attributes are 
introduced below.

%
%

\subsubsection{Security-related attributes.} The security challenge
aims to reduce the adversary's probability to successfully perform
a mafia, distance, or terrorist fraud attack\footnote{Another type of
fraud, named distance hijacking, was recently introduced by Cremers,
Rasmussen, Schmidt, and \v{C}apkun~\cite{CRS2012}, but this fraud is
usually disregarded in the analysis.}. The three following attributes
are consequently considered in this paper:

\begin{itemize}
\item \textit{Mafia fraud resistance ($\resm$).} Probability for an
  adversary to successfully perform a mafia fraud attack according to
  the Framework~\cite{ABKLM2011} already mentioned in
  Section~\ref{section:introduction}.
\item \textit{Distance fraud resistance ($\resd$).} Probability for an
  adversary to successfully perform a distance fraud attack according
  to the Framework.
\item \textit{Terrorist fraud resistance ($\rest$).} Probability for
  an adversary to successfully perform a terrorist fraud attack
  according to the Framework.
\end{itemize}

Other security-related attributes are the number of rounds $\nr$ and 
the size $\sm$ of the messages exchanged 
during the fast phase. On the one hand, most distance bounding 
protocols 
can arbitrarily increase $\sm$ while keeping $\nr$ constant~\cite{AFM2009}, 
which enhances their security. On the other hand, 
security can also be improved by simply 
increasing $\nr$. Both attributes are indeed related by the equation $\nbe = 
2\cdot\nr \cdot\sm$, where $\nbe$ represents the number of bits exchanged 
during 
the fast phase. We 
therefore consider $\nbe$ to be a security-related attribute that encompasses 
both $\nr$ and $\sm$. 

\begin{itemize}
\item \textit{Number of bits exchanged ($\nbe$).} Number of bits exchanged 
during the 
fast phase.
\end{itemize}

%


\subsubsection{Implementability-related attributes.}

When Hancke and Kuhn~\cite{HK2005} proposed a simple and lightweight design of 
distance bounding protocol, the objective was to reduce the number of 
cryptographic operations to be performed by the prover, and to avoid the use of 
a final slow phase. Consequently, these two implementability-related attributes 
are considered in this paper.
\begin{itemize}
\item \textit{Number of cryptographic operations performed by the prover 
($\nc$).} The number of cryptographic operations performed by the prover is 
considered, because it provides a preliminary technology-independent evaluation 
of the computational cost of the protocols.
\item \textit{Final slow phase ($\ns$).} Presence (or not) of a final slow 
phase in the protocol. 
\end{itemize}

Later on, memory usage became a concern as well, because the prover in
the tree-based protocol~\cite{AT2009} pre-computes a tree whose size
is exponential w.r.t.\ the number of rounds of the fast phase. Memory is
consequently considered as an implementability-related attribute:
\begin{itemize}
\item \textit{Memory used by the prover ($\nm$).} Maximum size of the volatile 
memory that the prover needs in order to store the values used during the 
protocol execution. Note that in practice, the actual size of the memory can be 
smaller because memory cells might be released and subsequently refilled with 
other values during the execution of the protocol. Considering the prover's 
memory instead of the verifier's memory is motivated by the prevailing design 
of distance bounding protocols where the prover needs to pre-compute all the 
possible answers before the fast phase, whereas verifying the prover's answers 
might not require heavy pre-computation and can be performed at the end of the 
protocol. 
\end{itemize}


Finally, the implementation complexity of a distance bounding protocol
strongly depends on the technology considered. This makes it
challenging to perform an objective evaluation with that respect. In
particular, some protocols require channels that carry atomic symbols
containing more than one bit of information. Although technologically
feasible, this requirement is strong enough to be taken into account
when comparing two protocols. In the same vein, some protocols use
multiple-bit exchanges during the fast phase, while a conservative
assumption since Desmedt {\etal}'s work~\cite{BD1990} consists in
considering 1-bit messages. This clear distinction between those distance 
bounding 
protocols that use single-bit exchanges during the fast phase and those that 
use multiple-bit exchanges is captured by the following 
implementability-related 
attribute.

\begin{itemize}
\item \textit{Multiple-bit exchange ($\nb$).} A binary attribute stating the 
use (or not) of multiple-bit exchanges. 
\end{itemize}

\subsection{Attribute spaces and (non)domination}

When solving decision-making problems, it is important to
consider a notion of \emph{approximate equality} on the domains of the
attributes. To illustrate
this, consider someone who wants to buy a second-hand car among cars
that differ on mileage and price only. When mileages are different but
very close, they can be considered in the same mileage range, and only
the price should lead the decision. 

In this section, we first provide the general terminology and notation, 
afterwards we define the attribute domain and approximate equality for every 
considered attribute.


\begin{definition}[Approximate equality]
\label{def:appeq}
Let $(\aset,<)$ be a totally ordered set.
An \emph{approximate equality relation} $\simil\colon \aset \times
\aset$ is a relation satisfying, for all $x,y,z \in \aset$,
\[
\begin{array}{c}
x \simil x\\
x \simil y \implies y \simil x\\
x \simil z \land x<y<z \implies x \simil y \land y \simil z\text{.}\\
\end{array}
\]
\end{definition}

The first two properties state that approximate equality satisfies
reflexivity and symmetry. The third property expresses that it is
consistent with the total order on $\aset$. Notice that approximate
equality is not an equivalence relation, because it doesn't satisfy
transitivity. The reason is that many small differences can add up to
a large difference.

Given a totally ordered set with approximate equality, we can define
the relation $\less\colon\aset\times\aset$ by
$$
x \less y \iff x<y \land x\not\simil y\text{.}
$$
Similarly, we define $x \lesssimil y$ by $x \less y \lor x \simil y$
and the symmetric cases $x \more y$
and $x \moresimil y$ by $y \less x$ and $y \lesssimil x$, respectively.
Next, we extend these comparison operators to attribute spaces.
\begin{definition}
Let $\indexset$ be an index set, then
a family of ordered sets with approximate equality
$(\aset_i,<_i,\simil_i)_{i\in\indexset}$ is called an \emph{attribute
space}.
\end{definition}

For an index set $\indexset = \{1, \ldots, n\}$, we simplify notation
by stating that 
$\aspace = \aset_1 \times \ldots \times \aset_n$ is an attribute space
and that its elements are of the form $\myvector{x}=(x_1, \ldots, x_n)$.
We define the \emph{dominant relation} $\less\colon\aspace\times\aspace$
for $\myvector{x},\myvector{y}\in\aspace$ by
$$\myvector{x} \less \myvector{y} \iff \forall i\in\indexset\, (x_i \lesssimil 
y_i)
  \land \exists i\in\indexset\, (x_i \less y_i)\text{.}$$
  
If $\myvector{x} \less \myvector{y}$, we say that $\myvector{x}$ 
\emph{dominates} $\myvector{y}$, otherwise, if $\myvector{x} \not\less
\myvector{y}$, we say that $\myvector{y}$ is \emph{nondominated} by 
$\myvector{x}$. Similarly, given $\asubset\subseteq\aspace$ 
and 
$\myvector{x}\in\asubset$, we say that 
$\myvector{x}$ is
\emph{nondominated} in $\asubset$ if
$$\neg\exists \myvector{y}\in\asubset\, (\myvector{y} \less 
\myvector{x})\text{.}$$


Given that we are considering eight different attributes, we next define a
totally ordered set with approximate equality relation for the eight 
considered attributes: $\resm, \resd, \rest, \nbe, \nc, \nm, \ns$, and $\nb$.

\begin{itemize}
	\item $(\aset_{i},<_{i},\simil_{i})_{i \in \{\resm, \resd, \rest\}}$: The 
	attributes related to the three types of fraud are in the probability domain
	$[0,1]$, \ie $\aset_{i} = [0,1]$ for $i \in \{\resm, \resd, \rest\}$. 
	In order to provide reasonable approximate equality relations for the three 
	probability-based attributes, we consider that the adversary's
  probability of success should be more refined as it approaches $0$.
  Therefore, a security value $x$ can be represented by the 
	interval $\left(\frac{x}{2}, 2x\right)$.
  The approximate equality relations are thus defined as follows.
$$\forall i \in \{\resm, \resd, \rest\} (x \simil_{i} y \iff \frac{x}{2} < y < 
2x).$$
The fact that this relation satisfies the three requirements from
Definition~\ref{def:appeq} follows from simple algebraic reasoning.
\item $(\aset_{i},<_{i},\simil_{i})_{i \in \{\nbe,\nc\}}$: Both the number of 
bits exchanged ($\nbe$) in the fast 
phase  and the number of 
cryptographic operations ($\nc$) are in the 
domain of the natural numbers 
$\mathbb{N}$. Their approximate equality relations $\simil_{\nc}$ and 
$\simil_{\nbe}$ are simply the equality in $\mathbb{N}$.
\item $(\aset_{\nm},<_{\nm},\simil_{\nm})$: Memory 
($\nm$) is in the domain of the natural numbers $\mathbb{N}$, and its 
approximate equality relation is defined by scaling from bits to kilobits. 
Defining $\simil_{\nm}$ in that way is a pragmatic approach 
based on experience in the field of contactless systems where any saving on a 
single kilobit is worthy. However, decision
makers could use a different relation, based on, e.g.,
megabytes. Formally, $\simil_{\nm}$ is defined as follows.
$$
x \simil_{\nm} y \iff  |x-y| < 1024\text{.}
$$
\item $(\aset_{\ns},<_{\ns},\simil_{\ns})$: Presence of a final slow phase 
($\ns$) is a nominal attribute in the Boolean domain. Protocols avoiding 
this phase are normally designed for low-cost devices~\cite{HK2005}. We thus 
define both the total order and the approximate equality relations as follows.
$$
x <_{\ns} y \iff  x = \texttt{false} \wedge y = \texttt{true}\text{.}
$$
$$
x \simil_{\ns} y \iff  x = y\text{.}
$$
\item $(\aset_{\nb},<_{\nb},\simil_{\nb})$: Use of a multiple-bit channel 
($\nb$) is also in the Boolean domain. A single-bit exchange protocol can be 
easily improved by transforming it to a multiple-bit protocol~\cite{AFM2009}. 
Consequently, we define the total order and the approximate 
equality relation as follows.
$$
x <_{\nb} y \iff  x = \texttt{false} \wedge y = \texttt{true}\text{.}
$$
$$
x \simil_{\nb} y \iff  x = y\text{.}
$$
\end{itemize}

\subsection{Solution}

The methodology introduced in this paper does not aim to identify the
best protocol in a general way but, instead, to identify the set of 
\emph{nondominated protocols}. Intuitively, a nondominated protocol satisfies 
that it is not possible to improve by moving away from it to another protocol 
without degrading the result w.r.t.\ at least one attribute. 

Providing a given protocol with attribute values typically 
requires one to specify values for protocol-specific parameters,
\emph{e.g.,} the 
number of rounds. We thus consider \emph{protocol instances}, which
are protocols for which all such parameters have been instantiated and
whose attribute values can be unambiguously 
determined. In order to not introduce additional notation, we simply represent 
this one-to-many relation from protocols to protocol instances by means of 
identifiers. In short, a \emph{protocol instance} is a pair $(\pname, x)$ where 
$\pname$ is an identifier that uniquely identifies a 
full-specification of a protocol, and $x \in \aspace$ provides the 
attribute values for the fully-specified protocol $PI$. We recall that $\aspace 
= \aset_{\resm} \times \aset_{\resd} \times
\aset_{\rest} \times 
\aset_{\nbe}  \times 
\aset_{\nc} \times \aset_{\nm} \times \aset_{\ns} \times \aset_{\nb}$ is the 
attribute space defined in Section~\ref{sec:methodology} over the index set 
$\indexset = \{\resm, \resd, \rest, \nbe,  
\nc, \nm, \ns, \nb\}$. 

%
%

\begin{definition}[Solution]\label{def:solution}
Given a set of protocol instances $\asubset$, a
solution in our 
methodology is the subset $\asolutions \subseteq \asubset$ of maximum 
cardinality such that for every $(P_x, \myvector{x}) \in \asolutions$ 
there does not exist $(P_y, \myvector{y}) \in \asubset$ such that $\myvector{y} 
\less \myvector{x}$. 
We say, in 
this case, that $(\pname, \myvector{x})$ is nondominated in 
$\asubset$. 
\end{definition}

We also say that a protocol 
instance $(P_x, \myvector{x})$ dominates another protocol instance $(P_y, 
\myvector{y})$ if and only if $\myvector{x} \less \myvector{y}$. If 
$\myvector{x} \not\less \myvector{y}$, we say that $(P_y, \myvector{y})$ 
is nondominated by $(P_x, \myvector{x})$.

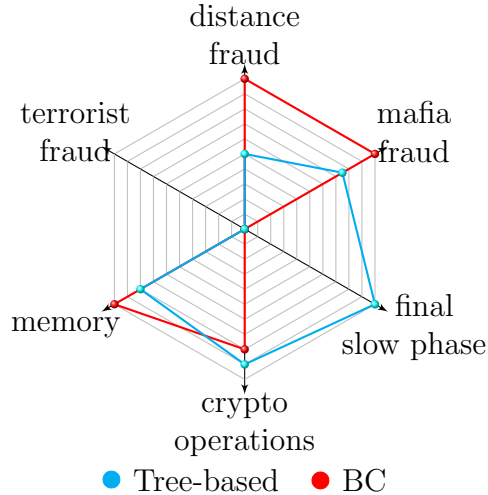
\begin{figure}[hbtp]\centering
\begin{tikzpicture}[label distance=.15cm,rotate=30,scale=0.4]
 \tkzKiviatDiagram[label distance=2cm,label space=1.5]{mafia fraud,distance 
 fraud,terrorist fraud,{memory\ \;},crypto operations,{\ \ final slow phase}}
  \tkzKiviatLine[thick,
                 color      = red,
                 mark       = ball,
                 ball color = red,
                 mark size  = 4pt,
                 opacity    = .2, 
                 fill=red!0](10,10,0,10,8,0)
 \tkzKiviatLine[thick,
                 color      = Cyan,
                 mark       = ball,
                 ball color = Cyan,
                 mark size  = 4pt,
                 opacity    = .2, 
                 fill=Cyan!0](7.5,5,0,8,9,10)
                 fill=blue!0](2,2,5,0,5,10)
                \filldraw[fill=Cyan, draw = Cyan] (-8,-5) 
                node[align=left, right] {\ Tree-based}
                circle [radius=8pt];
                \filldraw[fill=red, draw = red] (-2,-8.5) 
                node[align=left, 
                right] 
                {\ BC}
                circle [radius=8pt];                 
\end{tikzpicture}  
\caption{Spider chart for the protocol instances BC-$\{16\}$ and 
Tree-$\{16,8\}$.}\label{figure:dominated}
\end{figure}

To illustrate the nondominated relation between two protocol
instances, we make use of spider charts~\cite{spider_charts}.
Spider charts (also known under various other names, such as radar
charts) are simple graphs that make it possible to quickly compare the
relative scores of a number of alternatives along various axes.
An example of a spider chart is given in
Figure~\ref{figure:dominated}. There we 
present two protocol instances: one corresponds to Brands and Chaum's 
protocol~\cite{BC1993} (labeled ``BC-$\{16\}$'') when 
$n=16$, and the other one to the tree-based protocol~\cite{AT2009} (labeled 
``Tree-$\{16, 8\}$'') with depth 
equal to $8$ and $n=16$. The axes related to the types of fraud are 
logarithmically scaled from 1 (chart center) to $\log_2(\frac{1}{2^n})$ (chart 
outer); the axes related to the Boolean attributes are graduated with 
\texttt{true} (chart center) and \texttt{false} (chart outer). Finally, the 
axes concerning memory size and number of cryptographic operations are 
graduated 
from 10 (chart center) to 0 (chart outer).
In order to focus on the differences between the protocols, we will
often only display the attribute axes for which the protocols have
different values and the security-related attributes.
Consequently, in the current example, we omitted the $\nbe$ and $\nb$ axes.



A solution $S$, in the sense of Definition~\ref{def:solution}, can be seen as 
the set of relevant protocol instances a decision maker should focus on. A 
similar use can be given by distance bounding protocol designers, whose 
ultimately goal must be to include their protocols in $S$ w.r.t.\ some set of 
criteria. The role of $S$ is empirically illustrated in the next section where 
several state-of-the-art distance bounding protocols are evaluated and compared 
by applying our methodology.







\section{Methodology applied to current 
protocols}\label{sec:comparison}


This section reports on the results obtained after applying our
methodology to the protocols listed in
Table~\ref{table_protocol_list}. Instead of computing raw data to be
served as input to a state-of-the-art decision making tool, the
methodology has been implemented and published as an open source Java
project\footnote{The source code can be freely downloaded from
  \url{https://github.com/rolandotr/db_comparison}}. The computer tool, based 
  on Table~\ref{table_values}, 
comprises the thirteen distance bounding protocols listed in
Table~\ref{table_protocol_list} so as to generate protocol instances
as defined in Section~\ref{sec:methodology}. The decision to develop a computer 
tool is
supported by the growing number of distance bounding protocols
proposed and the continual refinement of their security
analysis~\cite{BBMSV2012,ABKLM2011,KAKSP2008}. Our tool therefore is aimed at 
facilitating the addition and modification of new protocols and criteria. 


\begin{table}
\caption{Formulas to compute the attribute values for every 
considered protocol. References to the sources of the formulas are 
included when applicable. Additional notation is introduced below.
\label{table_values}}
\begin{itemize}
	\item $\sigma$: size of signature, commitment, and MAC.
	\item $\delta$: size of the random nonces.
	\item $\ell$: depth of the tree in the tree-based 
	approach~\cite{AT2009}.
	\item $\alpha$: number of predefined challenges in KA~\cite{KA2011}.
	\item $p_f$: probability of occurrence of a void-challenge in 
	MP~\cite{MP2008}.
	\item $t$: size of the messages exchanged in the fast phase.
\end{itemize}  
\renewcommand{\arraystretch}{1.5}
{\setlength{\tabcolsep}{2pt}
\centering
\begin{tabular}{|c||c|c|c|c|c|c|c|}
\hline
\textbf{Protocols}&\textbf{$\resm$}&\textbf{$\resd$} 
&\textbf{$\rest$}&\textbf{$\ns$}&\textbf{$\nb$}&\textbf{$\nc$} &\textbf{$\nm$}\\
\hline
\hline
BC&$\left(\frac12\right)^n$~\cite{BC1993} 
&$\left(\frac12\right)^n$~\cite{BC1993} 
&$1$&\YES&\YES&$2$&$2n+3\sigma$\\
\hline
BB&$\left(\frac12\right)^n$~\cite{BB2005} 
&$\left(\frac12\right)^n$~\cite{BB2005} 
&$1$~\cite{ALM2011}&\YES&\YES&$4$&$3n+\delta$\\
\hline
MAD&$\left(\frac12\right)^n$~\cite{CBH2003} 
&$\left(\frac12\right)^n$~\cite{CBH2003} &$1$&\YES&\YES& 
$4$&$2n+2\delta+5\sigma$\\
\hline
HK&$\left(\frac34\right)^n$~\cite{HK2005} 
&$\left(\frac34\right)^n$~\cite{TMA2010}
 &$1$&\NO&\YES&$1$&$3n+2\delta$\\
\hline
MP & 
cf.~\cite{ABKLM2011}&cf.~\cite{ABKLM2011}&1&\YES&\YES&$2$& 
$4n+2\delta+\sigma$\\\hline
Swiss-Knife&$\left(\frac12\right)^n$~\cite{KAKSP2008}
&$\left(\frac34\right)^n$~\cite{KAKSP2008} 
&$\left(\frac34\right)^n$~\cite{KAKSP2008}&\YES&\YES&$2$&$3n+3\delta+2\sigma$\\
\hline
Tree-based 
&$\left(\left(\frac{1}{2}\right)^{\ell}\left(\frac{\ell}{2}+1\right)\right)^{\lfloor\frac{n}{\ell}\rfloor}$
& cf.~\cite{TMA2010} & $1$ & \NO &\YES&$1$ & 
$\left(2^{\ell+1}-1\right)\lfloor\frac{n}{\ell}\rfloor+$\\

&& &&&&&$2\delta+n$\\
\hline
Poulidor&cf.~\cite{TMA2010} 
& cf.~\cite{TMA2010} &$1$&\NO&\YES & 
$1$ & $5n+2\delta$\\
\hline
RC&$\left(\frac12\right)^n$~\cite{RC2010} 
&$\left(\frac12\right)^n$~\cite{RC2010} 
&$1$&\YES&\NO&$3$&$2\delta+2\sigma$\\\hline
YKHL&cf.~\cite{AK2013} 
& $\left(\frac{7}{8}\right)^n$ &$1$&\NO&\YES & 
$1$ & $5n+2\delta$\\
\hline
KA & 
cf.~\cite{KA2011}
 & 
$\left(\frac{3}{4}\right)^{n-\alpha}$~\cite{KA2011} & $1$ & \NO & \YES & 
$1$ & $4n+2\delta$\\
\hline
SKI& $\left(\frac{t+1}{2t}\right)^n$~\cite{BMV2013} & 
 $\leq \left(\frac{3}{4}\right)^n$~\cite{BMV2013}& 
 $\left(\frac{2t-2}{2t}\right)^n$~\cite{BMV2013} 
 &\NO&\NO&$1$&$n(t+1)+$\\
&& &&&&&$2\delta+2\sigma$\\
 
 \hline
TMA& cf.~\cite{TMA2014} & 
 cf.~\cite{TMA2014}& 
 1 &\NO&\NO&$1$&$4n+2\delta$\\\hline
\end{tabular}}
\end{table}


\subsection{Protocol instances}

Protocol instances are built by assigning values to 
protocol-specific parameters. In order to create a 
comprehensive set of protocol instances, we use ranges of 
values a bit wider than those considered in the 
literature. For instance, we consider protocols executing from $1$ to $256$ 
rounds during the fast phase, while in the literature this number varies from 
$16$ to $64$.  Other security-related parameters, namely the size of 
nonces 
($\delta$), secret keys 
($\kappa$), and cryptographic primitives ($\sigma$), are
considered to be large enough so that attacks based on, for example, short 
keys, are unfeasible. The remaining parameter values 
are detailed in Table~\ref{table_protocol_instances}. 

Once all parameter values are defined, we use 
Table~\ref{table_values} to obtain all 
protocol 
instances. This leads to a set $\asubset$ of $29184$ 
protocol 
instances, which is used as input to our 
methodology.

\begin{table}[h]
  \caption{Parameter values for the considered protocols. For the KA 
  protocol we use the 
  parameter $p_d$ instead of $\alpha$ given that 
  $\alpha = \lfloor p_d \times \nr 
  \rfloor$~\cite{KA2011}.\label{table_protocol_instances}}
  \centering
   \setlength{\tabcolsep}{12pt}
  \begin{tabular}{ @{} l  l  l  @{}}
    \toprule %
    \textbf{Protocol}        & \textbf{Identifier} & \textbf{Parameter values} 
    \\\cmidrule(l){1-3}
    BC & BC-$\{n\}$ & $n \in \{1, \cdots, 256\}$ \\\hline
    MAD & MAD-$\{n\}$ & $n \in \{1, \cdots, 256\}$ \\\hline
    BB & BB-$\{n\}$ & $n \in \{1, \cdots, 256\}$ \\\hline
    HK & HK-$\{n\}$ & $n \in \{1, \cdots, 256\}$ \\\hline
    \multirow{2}{*}{MP} & \multirow{2}{*}{MP-$\{n, p_f\}$} & $n \in \{1, 
    \cdots, 
    256\}$ \\
     & & $p_f \in \{0, 0.05, 0.01, \cdots, 1\}$\\\hline
    Swiss-Knife & Swiss-Knife-$\{n\}$ & $n \in \{1, \cdots, 256\}$ \\\hline
    \multirow{2}{*}{Tree-based} & \multirow{2}{*}{Tree-$\{n, \ell\}$} & $n \in 
    \{1, \cdots, 256\}$ \\
     & & $\ell \in \{1, 2, \cdots, 32\}$\\\hline
    Poulidor & Poulidor-$\{n\}$ & $n \in \{1, \cdots, 256\}$ \\\hline
    RC & RC-$\{n\}$ & $n \in \{1, \cdots, 256\}$ \\\hline
    YKHL & YKHL-$\{n\}$ & $n \in \{1, \cdots, 256\}$ \\\hline
    \multirow{2}{*}{KA} & \multirow{2}{*}{KA-$\{n, p_d\}$} & $n \in \{1, 
    \cdots, 256\}$ \\
     & & $p_d \in \{0, 0.05, 0.01, \cdots, 1\}$\\\hline
    \multirow{2}{*}{SKI} & \multirow{2}{*}{SKI-$\{n, t\}$} & $n \in \{1, 
    \cdots, 256\}$ \\
     & & $t \in \{2, 3, \cdots, 32\}$\\\hline
    TMA & TMA-$\{n\}$ & $n \in \{1, \cdots, 256\}$ \\
    \bottomrule
  \end{tabular}
\end{table}

\subsection{Comparison}




Comparing is definitely a decision making task. Decisions ought to be made for 
the sake of providing 
meaningful 
results. 
Nevertheless, the comparison problem 
differs from classical decision making 
problems in the role of the decision maker. 
The former problem should not reflect the point of view of 
the decision maker, but conciliate decisions and criteria 
based on a proper understanding of the problem and an 
exhaustive literature research. 

Along this article we have made a couple of 
decisions already. For instance, distance bounding 
protocols that fail on 
achieving any sort of 
authentication were discarded in 
Section~\ref{section:stateoftheart}.  
Section~\ref{sec:methodology} limits the 
number of considered attributes to $8$ by choosing those 
frequently used in the literature. And 
Section~\ref{sec:comparison} defines a wide range of 
parameter values in order to generate a comprehensive set 
of protocol instances. We claim that all these decisions 
are consistent with the state-of-the-art in distance 
bounding and, therefore, keep our experiments as fair as 
possible. 

Our last decision concerns a security criterion: 
mafia fraud resistance. All distance bounding protocols 
\emph{must} resist to mafia fraud to some extent. We thus 
consider 
different upper-bounds on the probability of success of an 
adversary mounting this type of fraud. More precisely, 
given the set $\asubset$ of $29184$ protocol instances 
defined 
previously and a probability value $y \in [0,1]$, we define 
the set $\asubset[y] = \{(\pname, 
x) 
\in \asubset | y \leq x_{\resm}\}$\footnote{We recall 
that $x$ is in 
	the attribute space $\aspace = \aset_{\resm} \times 
\aset_{\resd} \times
	\aset_{\rest} \times 
	\aset_{\nbe} \times 
	\aset_{\nc} \times \aset_{\nm} \times \aset_{\ns}\times 
\aset_{\nb} $ and 
	$x_i \in D_i$ for 
	every $i \in \{\resm, \resd, \rest, \nbe,  
	\nc, \nm, \ns, \nb\}$. 
} containing those protocols whose resistance to 
mafia fraud is bounded by $y$. In what follows we do not 
longer consider the whole set $\asubset$, but subsets 
$\asubset[y]$ for different values of $y$.

To illustrate further the 
need of this decision let us consider a protocol that does 
nothing. Because this protocol requires no resource to be 
implemented, it would be nondominated even though it can be hardly considered a 
distance bounding protocol. Considering 
$\asubset[y]$ for some $y < 1$ instead of $\asubset$, provides a quantifiable 
security guarantee in terms of mafia fraud that can only be provided by 
actual distance bounding protocols. Moreover, varying $y$ allows us 
to see how the set of nondominated protocols evolves when 
$y$ decreases. Table~\ref{tab_evolution} shows such 
evolution considering $y$ to range within the set 
$\{2^{-1}, 2^{-16}, 
2^{-32}, 
2^{-64}, 2^{-96}, 2^{-128}\}$. 

\begin{table}
\footnotesize  
\setlength\LTleft{-30pt}            
\setlength\LTright{-30pt}           
\caption{Nondominated protocol instances for different sets 
$\asubset[y]$. Every 
security value 
$p$ and memory value $m$ has been scaled according to the equations 
$2^{\lceil \log_2 p \rceil}$ and $\lfloor  m/1024 \rfloor$ 
respectively. For the sake of 
compactness, this table only shows for each protocol the 
nondominated 
protocol instance (if any) with fewer bits exchanged during 
the fast phase. The total 
number of nondominate protocols is given in the last 
column.
\label{tab_evolution}}
\centering
\setlength{\tabcolsep}{4pt}
\smallskip\noindent
\resizebox{\linewidth}{!}{%
\begin{tabular}{ @{} c | c | c c c c c c c c | c @{}}
\toprule
\multirow{2}*{$y$} &	
\multicolumn{1}{c}{Nondominated}	 & 
\multicolumn{7}{c}{\textbf{Attribute values}}	\\
 & Prot. Instances  & $n$ & $p_m$ & $p_d$ & $p_t$ & $b$ & 
 $c$ & $s$ 
 & $f$ & total\\ 
 \cmidrule(l){1-11}
\multirow{6}*{\ $2^{-1}$ \ }
& BC-\{1\} & $1$  & $2^{-1}$  & $2^{-1}$  & $2^{0}$  & $$\texttt{false}$$  & 
$2$  & $0\texttt{Kb}$  & $$\texttt{true}$$  & $256$\\ 
 & KA-\{2, 0.5\} & $2$  & $2^{-1}$  & $2^{-0}$  & $2^{0}$  & 
 $$\texttt{false}$$  & $1$  & $0\texttt{Kb}$  & $$\texttt{false}$$  & $10$\\ 
 & SKI-\{3, 2\} & $3$  & $2^{-1}$  & $2^{-1}$  & $2^{-3}$  & $$\texttt{true}$$  
 & $1$  & $1\texttt{Kb}$  & $$\texttt{false}$$  & $254$\\ 
 & SwissKnife-\{1\} & $1$  & $2^{-1}$  & $2^{-0}$  & $2^{-0}$  & 
 $$\texttt{false}$$  & $2$  & $1\texttt{Kb}$  & $$\texttt{true}$$  & $255$\\ 
 & TMA-\{2\} & $2$  & $2^{-1}$  & $2^{-1}$  & $2^{0}$  & $$\texttt{false}$$  & 
 $1$  & $0\texttt{Kb}$  & $$\texttt{false}$$  & $1$\\ 
 & Tree-\{2, 2\} & $2$  & $2^{-1}$  & $2^{-0}$  & $2^{0}$  & 
 $$\texttt{false}$$  & $1$  & $0\texttt{Kb}$  & $$\texttt{false}$$  & $400$\\ 
 \cmidrule(l){1-11}
\multirow{6}*{\ $2^{-16}$ \ }
 & BC-\{16\} & $16$  & $2^{-16}$  & $2^{-16}$  & $2^{0}$  & $$\texttt{false}$$  
 & $2$  & $0\texttt{Kb}$  & $$\texttt{true}$$  & $241$\\ 
 & KA-\{22, 0.55\} & $22$  & $2^{-16}$  & $2^{-4}$  & $2^{0}$  & 
 $$\texttt{false}$$  & $1$  & $0\texttt{Kb}$  & $$\texttt{false}$$  & $4$\\ 
 & Poulidor-\{23\} & $23$  & $2^{-16}$  & $2^{-8}$  & $2^{0}$  & 
 $$\texttt{false}$$  & $1$  & $0\texttt{Kb}$  & $$\texttt{false}$$  & $1$\\ 
& SKI-\{39, 2\} & $39$  & $2^{-16}$  & $2^{-16}$  & 
$2^{-39}$  & 
 $$\texttt{true}$$  & $1$  & $0\texttt{Kb}$  & $$\texttt{false}$$  & $218$\\ 
 & SwissKnife-\{16\} & $16$  & $2^{-16}$  & $2^{-6}$  & $2^{-6}$  & 
 $$\texttt{false}$$  & $2$  & $0\texttt{Kb}$  & $$\texttt{true}$$  & $241$\\ 
 & TMA-\{27\} & $27$  & $2^{-16}$  & $2^{-16}$  & $2^{0}$  & 
 $$\texttt{false}$$  & $1$  & $0\texttt{Kb}$  & $$\texttt{false}$$  & $1$\\ 
 & Tree-\{24, 6\} & $24$  & $2^{-16}$  & $2^{-10}$  & $2^{0}$  & 
 $$\texttt{false}$$  & $1$  & $0\texttt{Kb}$  & $$\texttt{false}$$  & $394$\\  
 \cmidrule(l){1-11}
\multirow{6}*{\ $2^{-32}$ \ }
 & BC-\{32\} & $32$  & $2^{-32}$  & $2^{-32}$  & $2^{0}$  & $$\texttt{false}$$  
 & $2$  & $0\texttt{Kb}$  & $$\texttt{true}$$  & $225$\\ 
 & KA-\{37, 0.85\} & $37$  & $2^{-32}$  & $2^{-2}$  & $2^{0}$  & 
 $$\texttt{false}$$  & $1$  & $0\texttt{Kb}$  & $$\texttt{false}$$  & $2$\\ 
 & Poulidor-\{42\} & $42$  & $2^{-32}$  & $2^{-16}$  & $2^{0}$  & 
 $$\texttt{false}$$  & $1$  & $0\texttt{Kb}$  & $$\texttt{false}$$  & $1$\\ 
 & SKI-\{78, 2\} & $78$  & $2^{-32}$  & $2^{-32}$  & $2^{-78}$  & 
 $$\texttt{true}$$  & $1$  & $0\texttt{Kb}$  & $$\texttt{false}$$  & $179$\\ 
 & SwissKnife-\{32\} & $32$  & $2^{-32}$  & $2^{-13}$  & $2^{-13}$  & 
 $$\texttt{false}$$  & $2$  & $0\texttt{Kb}$  & $$\texttt{true}$$  & $225$\\ 
 & TMA-\{53\} & $53$  & $2^{-32}$  & $2^{-32}$  & $2^{0}$  & 
 $$\texttt{false}$$  & $1$  & $0\texttt{Kb}$  & $$\texttt{false}$$  & $1$\\ 
 & Tree-\{48, 6\} & $48$  & $2^{-32}$  & $2^{-21}$  & $2^{0}$  & 
 $$\texttt{false}$$  & $1$  & $1\texttt{Kb}$  & $$\texttt{false}$$  & $368$\\  
 \cmidrule(l){1-11}
\multirow{6}*{\ $2^{-64}$ \ }
 & BC-\{64\} & $64$  & $2^{-64}$  & $2^{-64}$  & $2^{0}$  & $$\texttt{false}$$  
 & $2$  & $0\texttt{Kb}$  & $$\texttt{true}$$  & $193$\\ 
 & KA-\{73, 0.8\} & $73$  & $2^{-64}$  & $2^{-6}$  & $2^{0}$  & 
 $$\texttt{false}$$  & $1$  & $0\texttt{Kb}$  & $$\texttt{false}$$  & $4$\\ 
 & Poulidor-\{78\} & $78$  & $2^{-64}$  & $2^{-32}$  & $2^{0}$  & 
 $$\texttt{false}$$  & $1$  & $0\texttt{Kb}$  & $$\texttt{false}$$  & $1$\\ 
 & SKI-\{155, 2\} & $155$  & $2^{-64}$  & $2^{-64}$  & $2^{-155}$  & 
 $$\texttt{true}$$  & $1$  & $0\texttt{Kb}$  & $$\texttt{false}$$  & $102$\\ 
 & SwissKnife-\{64\} & $64$  & $2^{-64}$  & $2^{-26}$  & $2^{-26}$  & 
 $$\texttt{false}$$  & $2$  & $0\texttt{Kb}$  & $$\texttt{true}$$  & $193$\\ 
 & TMA-\{106\} & $106$  & $2^{-64}$  & $2^{-64}$  & $2^{0}$  & 
 $$\texttt{false}$$  & $1$  & $0\texttt{Kb}$  & $$\texttt{false}$$  & $1$\\ 
 & Tree-\{96, 6\} & $96$  & $2^{-64}$  & $2^{-43}$  & $2^{0}$  & 
 $$\texttt{false}$$  & $1$  & $2\texttt{Kb}$  & $$\texttt{false}$$  & $295$\\ 
 \cmidrule(l){1-11}
\multirow{6}*{\ $2^{-96}$ \ }
 & BC-\{96\} & $96$  & $2^{-96}$  & $2^{-96}$  & $2^{0}$  & $$\texttt{false}$$  
 & $2$  & $0\texttt{Kb}$  & $$\texttt{true}$$  & $161$\\ 
 & KA-\{113, 0.75\} & $113$  & $2^{-96}$  & $2^{-12}$  & $2^{0}$  & 
 $$\texttt{false}$$  & $1$  & $0\texttt{Kb}$  & $$\texttt{false}$$  & $5$\\ 
 & Poulidor-\{114\} & $114$  & $2^{-96}$  & $2^{-49}$  & $2^{0}$  & 
 $$\texttt{false}$$  & $1$  & $0\texttt{Kb}$  & $$\texttt{false}$$  & $1$\\ 
 & SKI-\{232, 2\} & $232$  & $2^{-96}$  & $2^{-96}$  & $2^{-232}$  & 
 $$\texttt{true}$$  & $1$  & $1\texttt{Kb}$  & $$\texttt{false}$$  & $25$\\ 
 & SwissKnife-\{96\} & $96$  & $2^{-96}$  & $2^{-39}$  & $2^{-39}$  & 
 $$\texttt{false}$$  & $2$  & $1\texttt{Kb}$  & $$\texttt{true}$$  & $161$\\ 
 & TMA-\{158\} & $158$  & $2^{-96}$  & $2^{-96}$  & $2^{0}$  & 
 $$\texttt{false}$$  & $1$  & $0\texttt{Kb}$  & $$\texttt{false}$$  & $1$\\ 
 & Tree-\{144, 6\} & $144$  & $2^{-96}$  & $2^{-64}$  & $2^{0}$  & 
 $$\texttt{false}$$  & $1$  & $3\texttt{Kb}$  & $$\texttt{false}$$  & $223$\\ 
 \cmidrule(l){1-11}
\multirow{6}*{\ $2^{-128}$ \ }
 & BC-\{128\} & $128$  & $2^{-128}$  & $2^{-128}$  & $2^{0}$  & 
 $$\texttt{false}$$  & $2$  & $0\texttt{Kb}$  & $$\texttt{true}$$  & $129$\\ 
 & KA-\{145, 0.8\} & $145$  & $2^{-128}$  & $2^{-12}$  & $2^{0}$  & 
 $$\texttt{false}$$  & $1$  & $0\texttt{Kb}$  & $$\texttt{false}$$  & $4$\\ 
 & Poulidor-\{148\} & $148$  & $2^{-128}$  & $2^{-65}$  & $2^{0}$  & 
 $$\texttt{false}$$  & $1$  & $0\texttt{Kb}$  & $$\texttt{false}$$  & $1$\\ 
 & SKI-\{219, 3\} & $219$  & $2^{-128}$  & $2^{-90}$  & $2^{-128}$  & 
 $$\texttt{true}$$  & $1$  & $1\texttt{Kb}$  & $$\texttt{false}$$  & $1$\\ 
 & SwissKnife-\{128\} & $128$  & $2^{-128}$  & $2^{-53}$  & $2^{-53}$  & 
 $$\texttt{false}$$  & $2$  & $1\texttt{Kb}$  & $$\texttt{true}$$  & $129$\\ 
 & TMA-\{210\} & $210$  & $2^{-128}$  & $2^{-128}$  & $2^{0}$  & 
 $$\texttt{false}$$  & $1$  & $1\texttt{Kb}$  & $$\texttt{false}$$  & $1$\\ 
 & Tree-\{160, 16\} & $160$  & $2^{-128}$  & $2^{-77}$  & $2^{0}$  & 
 $$\texttt{false}$$  & $1$  & $1280\texttt{Kb}$  & $$\texttt{false}$$  & 
 $150$\\ 

  \bottomrule
\end{tabular}}
\end{table}

According 
to Table~\ref{tab_evolution}, seven out of the thirteen considered protocols 
have at least 
one instance that is nondominated for some set 
$\asubset[y]$. In this case, we say that these protocols 
are nondominated. The seven nondominated
protocols are 
BC, KA, SKI, Swiss-Knife, TMA, Poulidor, and Tree-based. We intuitively explain 
this result as follows.

\begin{figure*}
\centering
  \subfigure[]{
  	\includegraphics{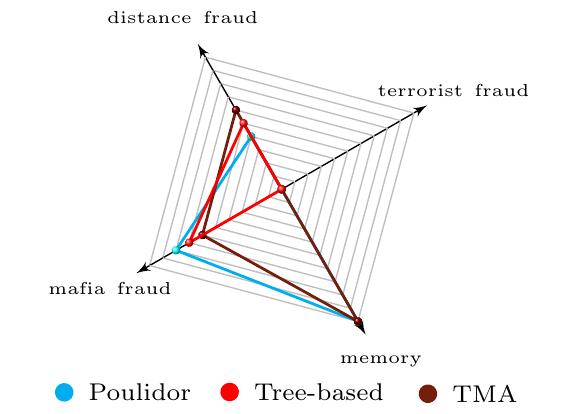}
  	\label{figure:dominated_tree_poulidor_tma}
  }
  \subfigure[]{
	\includegraphics{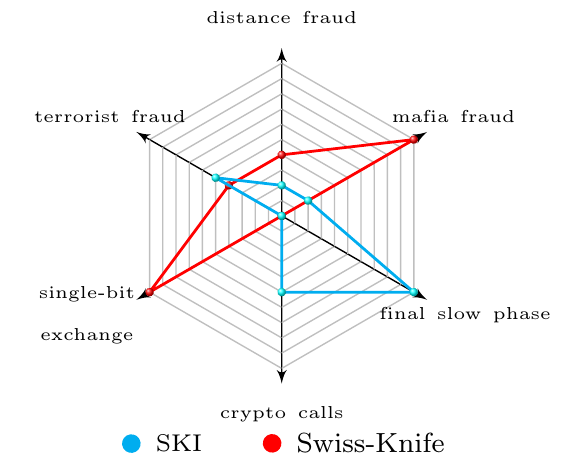}
	\label{figure:dominated_swiss_ski}
  }
  \caption{Two spider charts showing nondominated protocol instances. 
  Figure~\ref{figure:dominated_tree_poulidor_tma} considers the protocol 
  instances  Tree-based-$\{128\}$,  Poulidor-$\{128\}$, and TMA-$\{128\}$. 
  Figure~\ref{figure:dominated_swiss_ski} considers the protocol instances 
  Swiss-Knife-$\{128\}$ and SKI-$\{64,2\}$.  All axes 
  have been normalized with respect to 
  an ideal protocol instance executing $128$ rounds that 
  takes 
  the optimal value for each attribute.}
\end{figure*}

\begin{itemize}
\item BC, BB, MAD, and RC, achieve the 
optimal 
security
  in terms of both mafia and distance fraud (see
  Figures~\ref{Fig:mafia} and~\ref{Fig:distance} in the
  Appendix). Consequently, none of them can be dominated by 
  any of the
  remaining nine protocols. However, BC leaves out BB, MAD, 
  and RC, from the 
  set of nondominated protocols 
  because it requires fewer calls to cryptographic 
  functions. 
\item Swiss-Knife and SKI are the only protocols that 
resist to terrorist fraud 
  (see Figure~\ref{Fig:terrorist} in the Appendix). They do not dominate each 
  other as it is illustrated 
  by the Spider
  Chart~\ref{figure:dominated_swiss_ski}. Both 
  are thus nondominated.
\item Tree-based, Poulidor, and TMA, are the best in
  terms of distance fraud (see Figure~\ref{Fig:distance}) among the
  protocols using single-bit exchanges and a single cryptographic
  operation. Because they do not dominate
    each other (see the Spider
  Chart~\ref{figure:dominated_tree_poulidor_tma}), the 
  three 
  are included in the set of nondominated protocols. 
\item KA does not perform well in terms of 
distance fraud (see Figure~\ref{Fig:distance}). However, 
its resistance to mafia fraud can be as good as the one 
provided by the Tree-based protocol without demanding an 
exponential amount of memory (see 
Figure~\ref{Fig:mafia}). Therefore, KA is also nondominated. 
\end{itemize}

It is worth remarking that, according to 
Table~\ref{tab_evolution}, the 
set of nondominated protocols is rather stable 
with $y$. The only exception is Poulidor, that becomes a 
member of the set of 
nondominated protocols for $y \leq 2^{-16}$. 
This behavior is likely to be due to the fact that the actual 
distance fraud resistance 
of the Poulidor protocol cannot be 
computed yet~\cite{TMA2010,T2013}, but an upper-bound only.

\section{Conclusion}

In this article, we have proposed a methodology to evaluate and
compare distance bounding protocols. The methodology benefits from
experiences in the decision making field, and defines the most
relevant attributes that ought to be considered in terms of security
and implementability. An open-source computer software implementing
our methodology has been released, which supported the evaluation and
comparison of thirteen state-of-the-art distance bounding
protocols. Among the evaluated protocols, only seven are relevant
(nondominated) in terms of the considered criteria, namely resistance
to mafia, distance, and terrorist fraud, number of cryptographic
operations, memory, presence of a final slow phase, and use of a
multiple-bit channel. Clearly, most disqualified protocols had an
important role in the evolution of distance bounding protocols, but
they are obsolete today. Future designs of distance bounding protocols
must, therefore, prove to be nondominanted with respect to a set of
relevant criteria. Our results also show that the asymptotic analysis
of distance bounding protocols, as done commonly in the literature, is
inadequate and misleading. Finally, a clear side effect of our methodology is 
that it 
can be used for ad-hoc decision making where the decision maker is free to 
prioritize some attributes over others. 

\bibliographystyle{abbrv}

\section*{Appendix}

\begin{figure}[h]
\centering
  \includegraphics[angle = 270, width=0.9\textwidth]{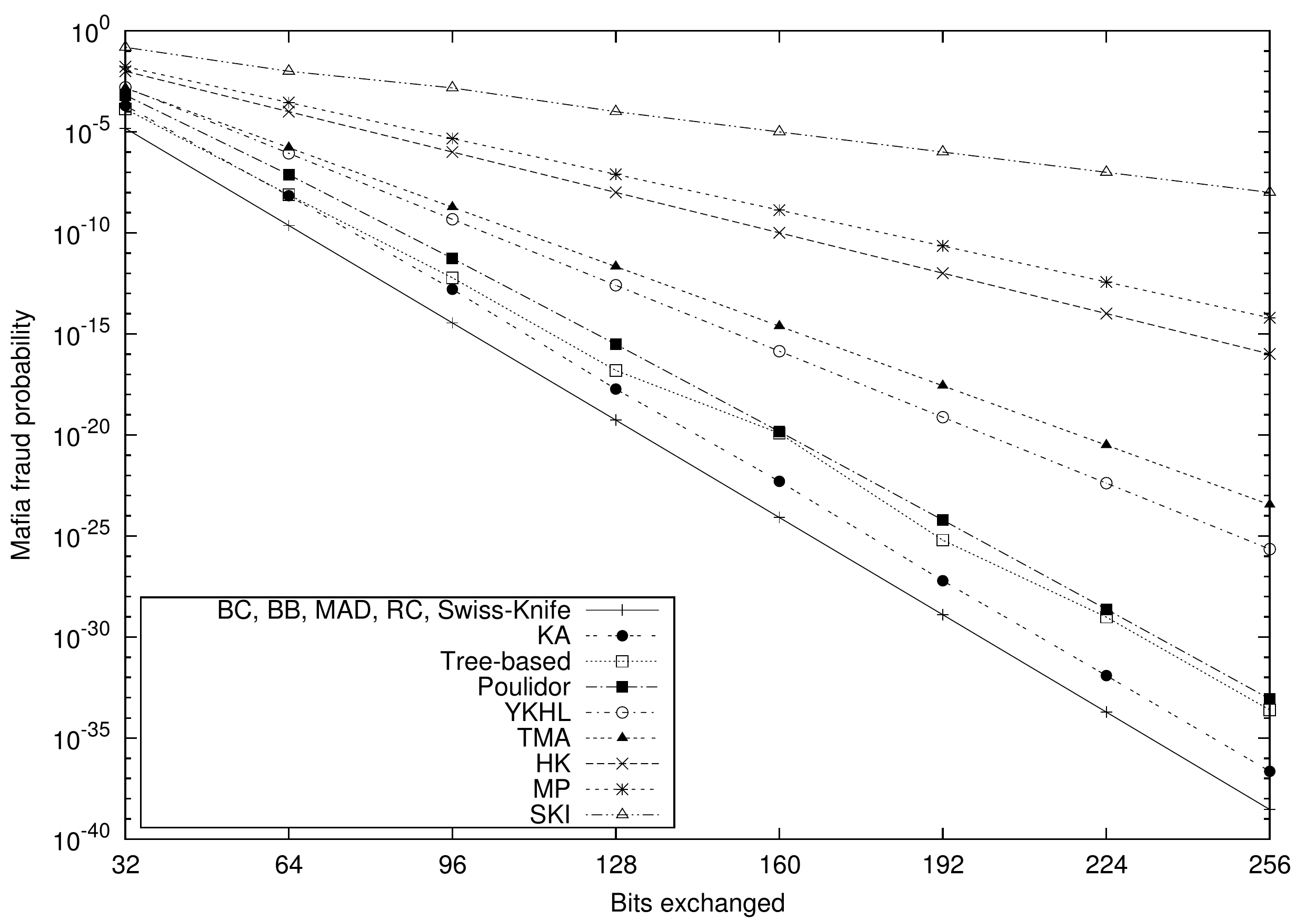}
   \caption{Mafia fraud resistance of the considered protocols 
      for $\nbe \in \{32, 64, \cdots, 258\}$.}\label{Fig:mafia}
\end{figure}

\begin{figure}
\centering
  \includegraphics[angle = 270, width=0.9\textwidth]{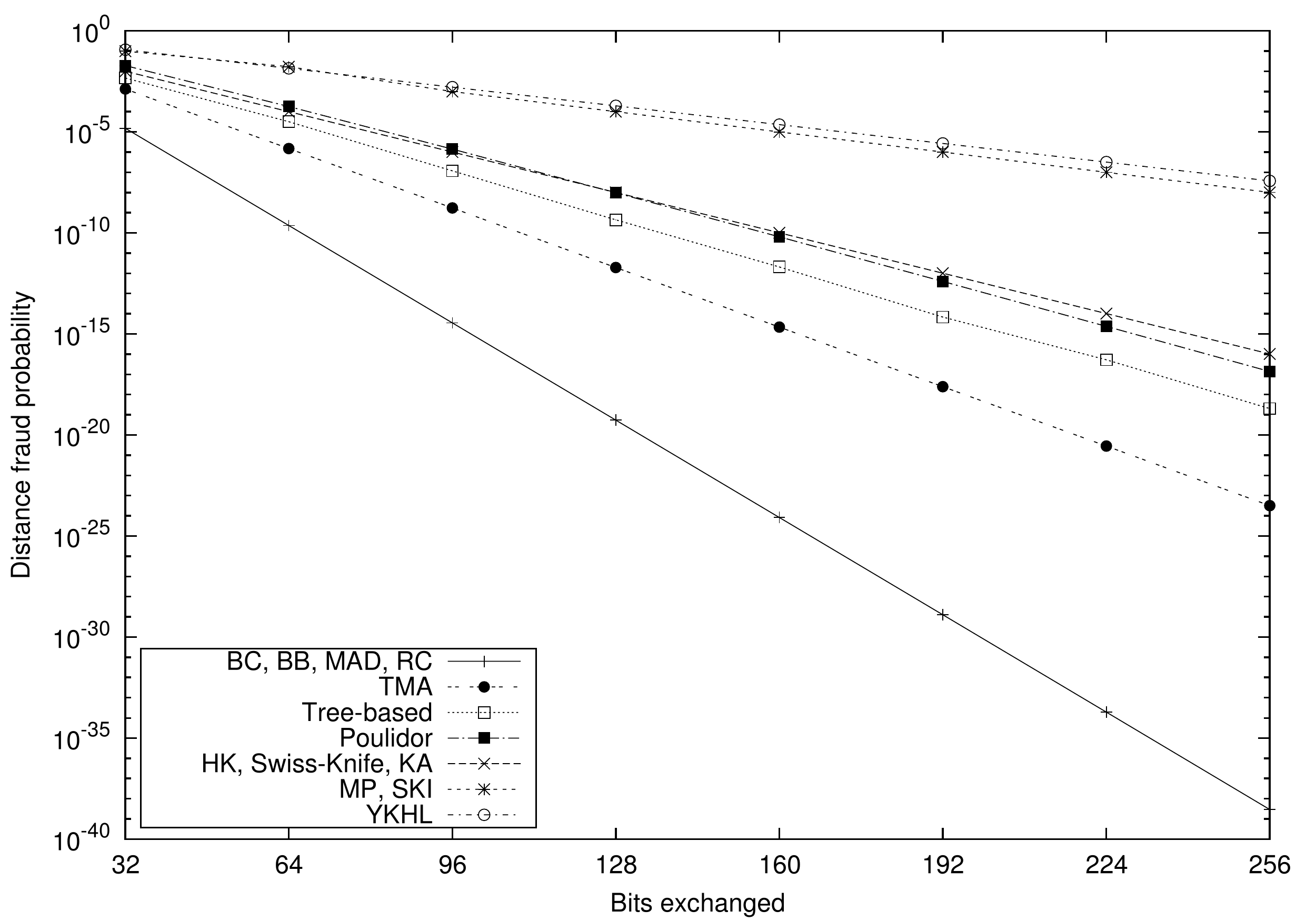}
   \caption{Distance fraud resistance of the considered protocols 
         for $\nbe \in \{32, 64, \cdots, 258\}$.}\label{Fig:distance}
\end{figure}

\begin{figure}
\centering
  \includegraphics[angle = 270, width=0.9\textwidth]{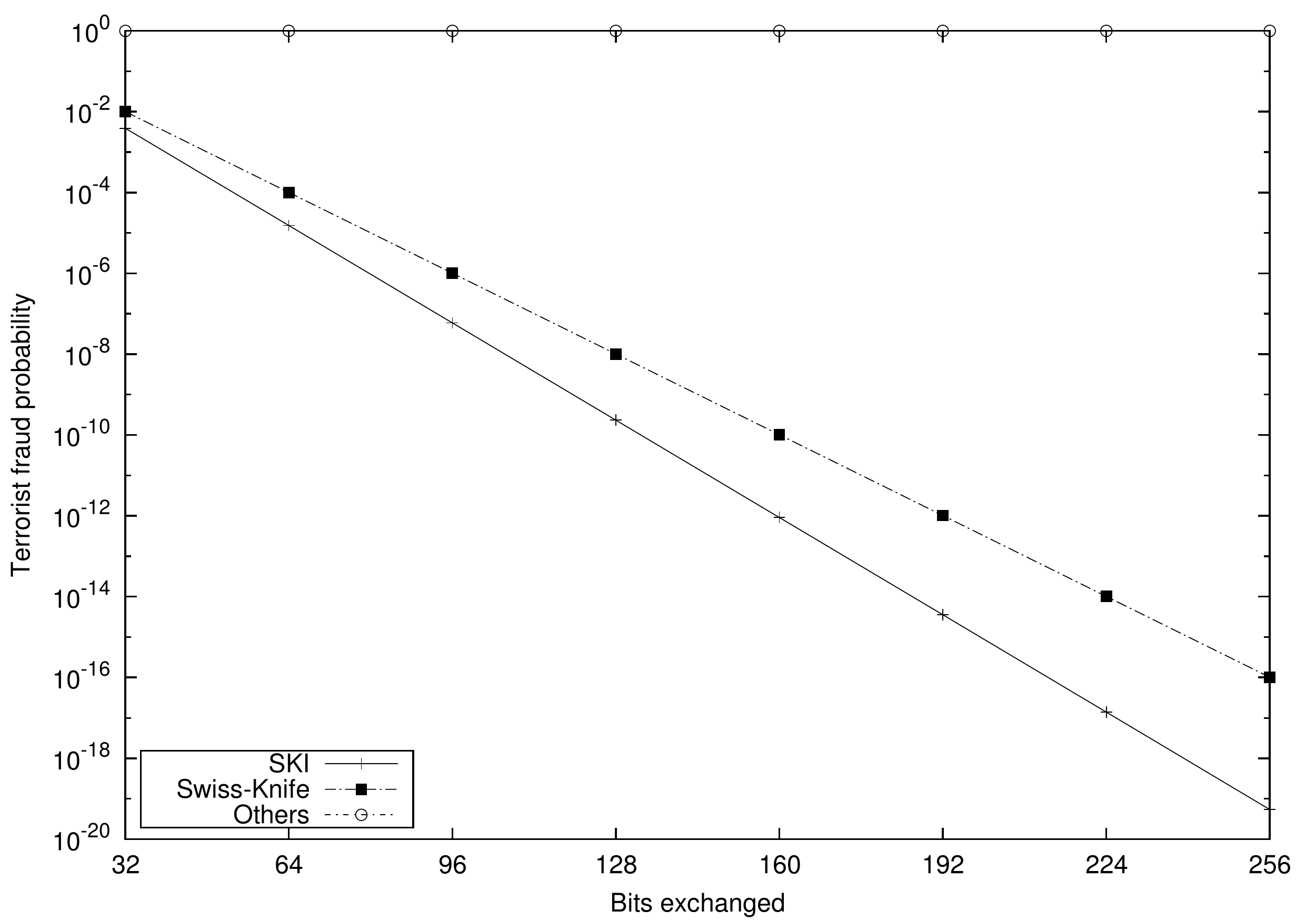}
   \caption{Terrorist fraud resistance of the considered protocols 
         for $\nbe \in \{32, 64, \cdots, 258\}$.}\label{Fig:terrorist}
\end{figure}

Figures~\ref{Fig:mafia},~\ref{Fig:distance}, and~\ref{Fig:terrorist}, depict 
the resistance of each protocol to mafia, distance, and 
terrorist fraud respectively. 
The 
attribute value for each fraud come from the protocol 
instance that minimizes 
it. For example, given $e = 32$, the resistance to mafia 
fraud of the KA 
protocol is taken from the protocol instance KA-$\{32, 
1\}$. On the contrary, its distance 
fraud resistance considering again $e = 32$ is taken from 
the protocol instance KA-$\{32, 0\}$. 

\end{document}